\documentclass[a4paper,11pt,parskip=full]{article}
\usepackage{pos}
\usepackage{aas_macros}
\usepackage{hyperref}
\usepackage{wrapfig}
\usepackage{xspace}
\usepackage{enumitem}

\newcommand{\g}{\ensuremath{\gamma}\xspace}
\newcommand{\agam}{\textit{newASTROGAM}\xspace}

\title{newASTROGAM -- The New MeV to GeV Gamma-ray Observatory}
\ShortTitle{newASTROGAM -- closing the MeV gap}

\author*[a,b]{D.~Berge}
\author[c]{M.~N.~Mazziotta}
\author[d]{M.~Tavani}
\author[e]{V.~Tatischeff}
\author[f]{U.~Oberlack}
\onbehalf{on behalf of the newASTROGAM proposal team}

\affiliation[a]{Deutsches Elektronen-Synchrotron DESY, Platanenallee 6, D-15738 Zeuthen, Germany}

\affiliation[b]{Humboldt-University of Berlin, Institute of Physics,
  Newtonstra{\ss}e 15, D-12489 Berlin, Germany}

\affiliation[c]{Istituto Nazionale di Fisica Nucleare, Sezione di
  Bari, via Orabona 4, I-70126 Bari, Italy} 
\affiliation[d]{INAF-IAPS Roma, via del Fosso del Cavaliere 100,
  I-00133 Roma, Italy} 
\affiliation[e]{Universit{\'e} Paris-Saclay, CNRS/IN2P3, IJCLab, F-91405
  Orsay, France}
\affiliation[f]{Institut für Physik \& Exzellenzcluster PRISMA+,
  Johannes Gutenberg-Universität Mainz, D-55099 Mainz, Germany}

\emailAdd{david.berge@desy.de}

\abstract{newASTROGAM is a breakthrough mission concept for the study
  of the non-thermal Universe from space with \g-rays in the energy
  range from 15~keV to 3~GeV. It is based on advanced space-proven
  detector technologies, which will achieve unprecedented sensitivity,
  angular and energy resolution combined with polarimetric
  capability. Since the MeV \g-ray energy range is the most
  under-explored electromagnetic window to the Universe, a mission in
  this energy range can for the first time sensitively address
  fundamental astrophysics questions connected to the physics of
  compact objects and merger events, jets and their environments,
  supernovae and the origin of the elements, potentially constrain the
  nature of dark matter and many more science objectives. The mission
  will detect and follow-up many of the key sources of multi-messenger
  astronomy in the 2040s.

  newASTROGAM provides an unprecedentedly broad energy coverage from
  keV to GeV energies. They payload concept consists of a Silicon
  tracker combined with a crystal calorimeter. Both detectors are
  surrounded by an anti-coincidence detector to reject charged cosmic
  rays. In addition, a thin X-ray coded mask provides very good
  imaging capabilities. Such a mission can uniquely detect \g-rays
  via the photoelectric effect, Compton scattering and
  electron-positron pair production. newASTROGAM is
  proposed to the ESA call for medium-class mission ideas (M8).}

\ConferenceLogo{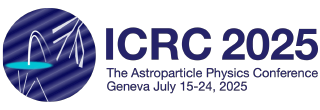}

\FullConference{39th International Cosmic Ray Conference (ICRC2025)\\
 15–24 July 2025\\
Geneva, Switzerland\\}

\begin{document}
\maketitle
\setlength{\parindent}{0pt}
\section{Introduction}
The \g-ray sky at MeV and GeV energies provides unique and crucial
information on the most energetic and intriguing phenomena of our
Universe.  Photons in this range provide fundamental insights into the
physics of nucleosynthesis, accretion, particle acceleration, strong
magnetic fields, and extreme-gravity
environments~\cite{2018JHEAp..19....1D}. Observing the sky at these energies with a large
field of view (FoV) and unprecedentedly high sensitivity is of
paramount importance in the era of multi-messenger astronomy, as
clearly underlined in the
\href{https://www.cosmos.esa.int/web/voyage-2050}{Voyage~2050} program
and
\href{https://www.nationalacademies.org/our-work/decadal-survey-on-astronomy-and-astrophysics-2020-astro2020}{Astro2020
  Decadal Survey} report.

\agam\footnote{\textbf{Mission homepage:} \url{https://www.new-astrogam.eu}} is designed as a unique multi-purpose observatory with scanning
and pointing capability, enabling imaging, polarimetry, line and
continuum spectroscopy, sub-millisecond timing, and fast transient
detection in the MeV-GeV energy domain. It will be the first
instrument with adequate sensitivity in the MeV range and simultaneous
GeV coverage, thus ensuring the broad-band information needed to
unravel many fundamental physical processes. \agam is {\it new}
because a hard-X-ray imager with arcminute source localization
capability has been added to the original design. This addition
broadens the purpose of the mission, which will not only complement
data from, but also provide trigger information to, prominent
facilities such as SKA, ALMA, Roman Space Telescope, ELT, newAthena,
CTAO, SWGO, LISA, the Einstein Telescope, IceCube-Gen2, KM3NeT.

While the mission will contribute to many science topics of interest
for a broad and diverse community, \agam promises major breakthroughs
in the following areas: 
\begin{itemize}[noitemsep]
\item physics of extreme cosmic accelerators from within the Galaxy to Cosmic Dawn;
\item nucleosynthesis, supernova explosions, chemical evolution of the Interstellar Medium (ISM);
\item Cosmic Ray (CR) sources and feedback;
\item the extreme Universe in the era of multi-messenger astronomy;
\item fundamental physics.
\end{itemize}

\section{Science}
\begin{figure}[htbp]
\centering
\includegraphics[width=0.8\linewidth]{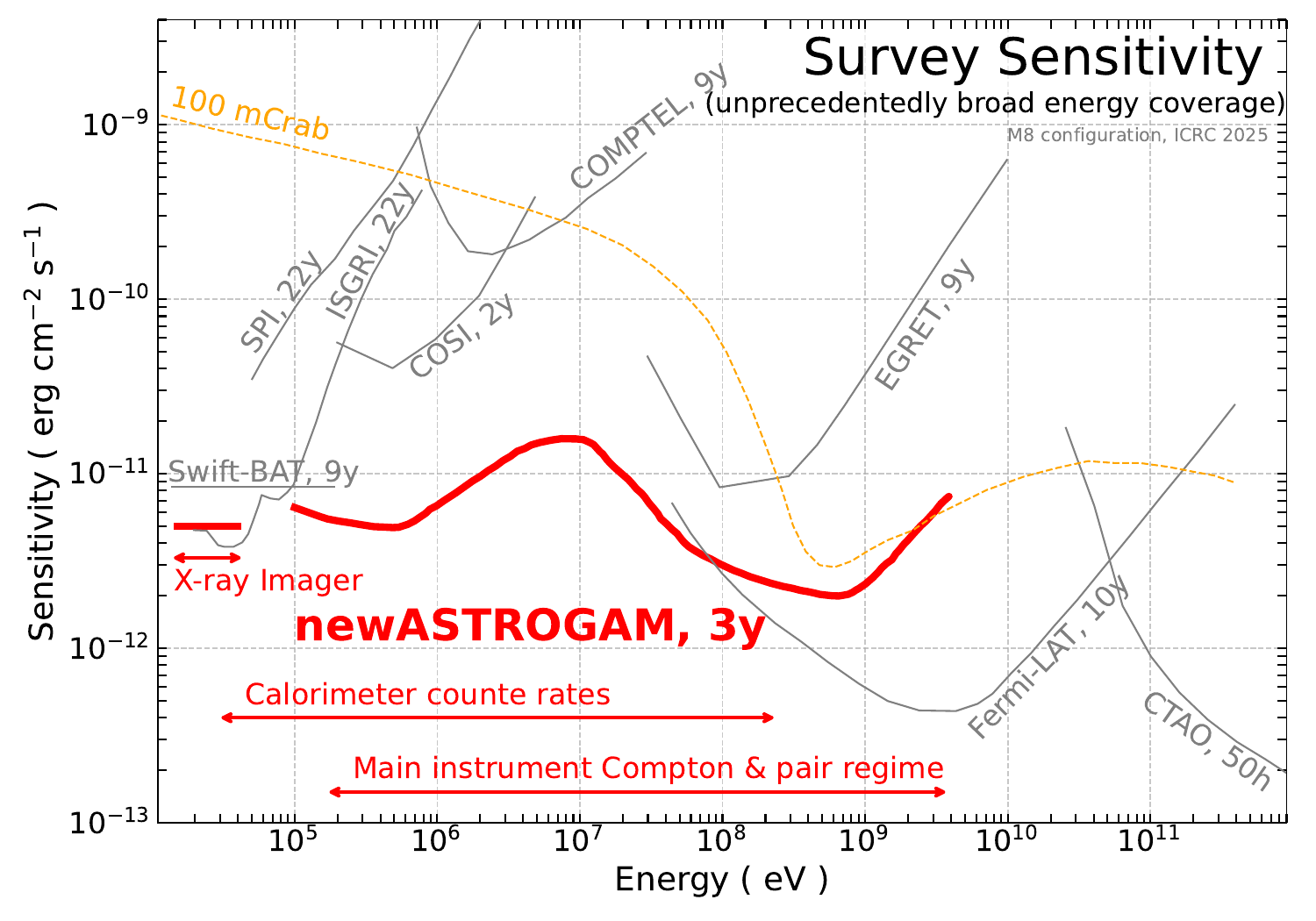}
\caption{Point-source (extra-galactic) continuum sensitivity of different X- and $\gamma$-ray instruments.
\label{fig:sensitivity}}
\end{figure}

With its large FoV ($\sim 1/3$ of the sky), all-sky scanning mode, and
arcmin localization capability, \agam will study the activity of
thousands of cosmic accelerators such as pulsars, stellar and
supermassive black holes (SMBHs), \g-ray bursts (GRBs), magnetars,
supernovae (SNe), and kilonovae.  With its unprecedented sensitivity
(see Fig.~\ref{fig:sensitivity}), it will provide the deepest all-sky
mapping of the MeV emission produced by the radioactive decay of newly
synthesized nuclei, CR interactions, $e^\pm$ pairs, and decisively
search the MeV--GeV sky for dark-matter (DM) signatures.  Finally, it
will help identify electromagnetic counterparts to gravitational-wave
(GW) and high-energy neutrino sources, expanding the pioneering role
of the \emph{Fermi} satellite.

\vspace{0.4cm}
\textbf{Insights into Extreme Acceleration Processes}
\vspace{0.1cm}

MeV-GeV variability studies of relativistic jets can be conducted with
\agam on timescales of seconds to hours for GRBs and from minutes to
years for AGN. For AGN jets they will provide crucial information on
the size and location of the acceleration site, and on the Lorentz
factor and any precession of the jet. The detailed shape of the
emission spectrum around the MeV peak places important constraints on
acceleration and radiation mechanisms, and maximum particle
energy~\cite{2025A&A...694L...3T}. \agam
data will also constrain the lepton/baryon content and magnetization
of the jets, whose knowledge is fundamental to assessing their
energetics, the microphysics of particle acceleration, and the impact
on their surroundings.  Analogous constraints on the sites and
mechanisms of particle acceleration will be placed on black-hole (BH)
and neutron-star (NS) X-ray binaries, recently probed by INTEGRAL and
{\it Fermi} as high-energy emitters, with the MeV band providing the
missing
link~\cite{2022ASSL..465..157P,2021NewAR..9301618M}.

\g-ray
polarimetry~\cite{2019NewAR..8701537G}
is a young field with a large discovery potential.  Polarisation
changes between X-rays and \g-rays can probe the location and
evolution of relativistic particle outflows.  \agam will enable
polarimetric measurements for tens of Galactic sources with
sensitivity to variations in time.  For accreting BHs, measurements of
the polarisation and the MeV cutoff energy can firmly establish the
nature of the observed MeV tail, disentangling its origin between the
jet and the disk corona.

Current observations of TeV-to-PeV neutrinos suggest the existence of
both a Galactic and an extragalactic population of hadronic PeVatrons
(PeV ion accelerators) mostly hidden to GeV-TeV \g-ray telescopes due
to strong \g-ray absorption in the source. A handful of temporal
associations between neutrinos and \g-ray flares suggest AGN jets and
Galactic \g-ray binaries as promising
PeVatrons~\cite{2021ApJ...921L..10B}.
Some steady neutrino
sources~\cite{2022Sci...378..538I}
are claimed to be associated with AGN coronae.  The accompanying
cascade radiation would preferentially appear in the MeV
band~\cite{2022ApJ...941L..17M},
making \agam's role decisive to decipher the processes of PeV particle
acceleration, both in Seyfert
galaxies~\cite{2024ApJ...961L..34M}
and in
blazars~\cite{2023ApJ...958L...2F}.

\agam will shed light on BH activity in the early Universe.  The most
luminous AGN, powered by accretion onto a SMBH of about
$10^9\,M_\odot$, are most abundant at redshifts
$z \gtrsim
3$~\cite{2022ApJ...940...77M},
before the cosmic noon in star-formation history. The jet emission
peaks in the MeV-GeV band, and largely dominates the radiative output
when the jet points in our direction (blazar). \agam can characterise
the jet activity of hundreds of objects, both in quiescence and during
flares. MeV blazars can be detected up to redshift
$z \sim
6-7$~\cite{2022ApJ...940...77M},
though most of the detected sources will have redshifts $z < 4$. The
opportunity is unique to shed light on the physics of this source
class, the cosmological evolution of SMBHs, the environment of their
growth, and the extent to which AGN jets influence galaxy
evolution~\cite{2021Galax...9...23S}.
This topic is also identified as a Voyage~2050 priority.

\vspace{0.4 cm}
\textbf{Explosive Nucleosynthesis and Chemical Evolution of Galaxies}
\vspace{0.1cm}

\begin{figure}[htbp]
\centering
\includegraphics[width=\linewidth]{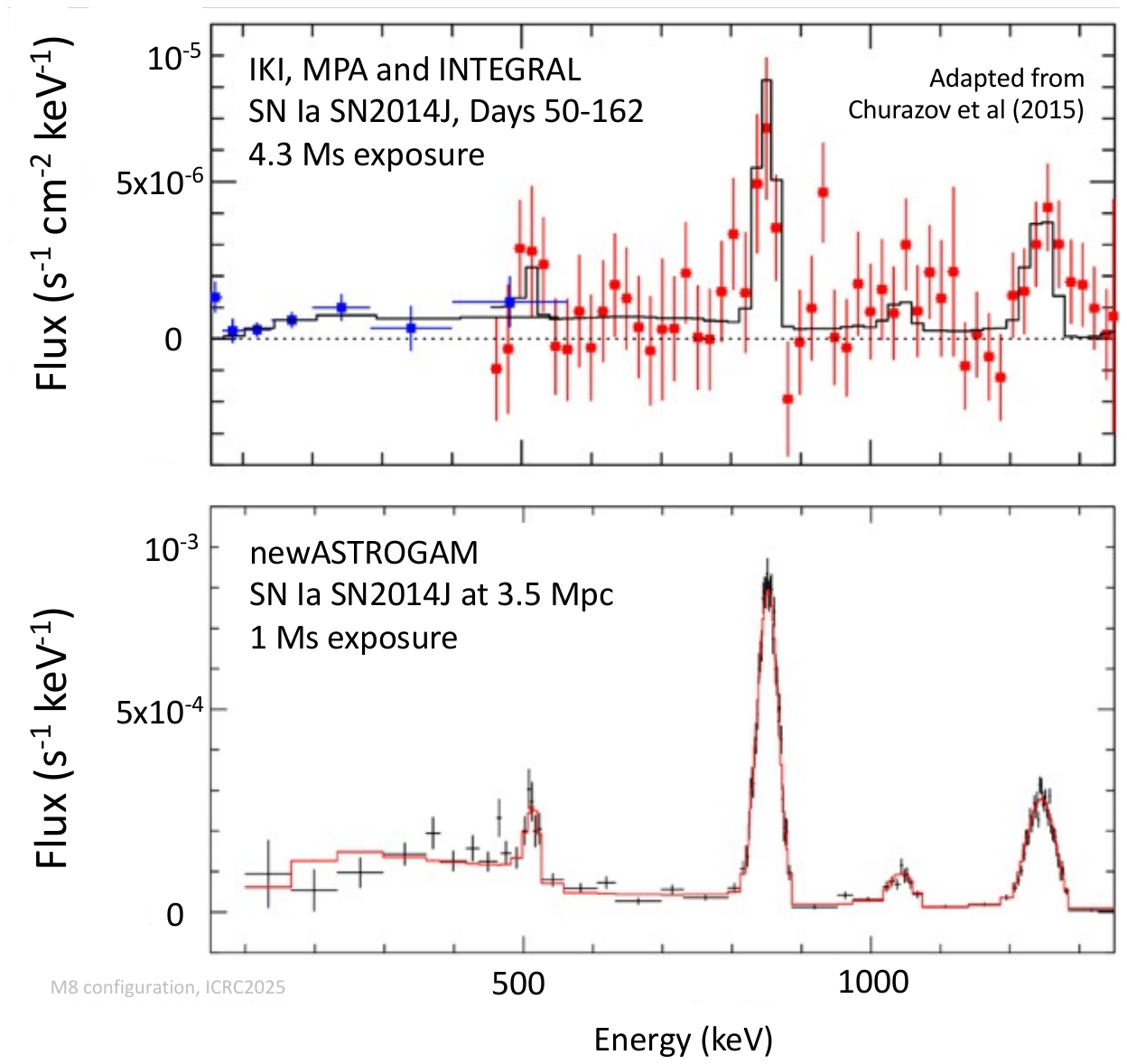}
\caption{Improvement over INTEGRAL (top,~\cite{2015ApJ...812...62C}) for a type-Ia SN. 
Lines from the $^{56}\mathrm{Ni}$ decay chain will clearly be resolved (bottom), yielding
the radioactive and total ejecta masses, a characteristic of the progenitors. 
\label{fig:SNIa_spec}}
\end{figure}

Interstellar enrichment in heavy nuclei is driven by
asymptotic-giant-branch stars, core-collapse supernovae (SNe),
white-dwarf explosions, and NS mergers (kilonovae), but their relative
contributions are not established at
all~\cite{2020RSPTA.37890301J}.
\agam can detect important nuclear decay lines with unprecedented
sensitivity and, for nominal mission lifetimes, improve over the
upcoming COSI
mission~\cite{2024icrc.confE.745T}
by factors 2.5 to 30, depending on the line.

The large FoV will help detecting \g-rays, the earliest photons to
escape explosion sites, before the optical peak, for optimal
constraints on the nature and mass of the progenitor and on the
explosion
itself~\cite{2021NewAR..9201606I}.
The detection of type-Ia SNe out to a distance of 30\, Mpc will enable
precise measurements of the total mass of $^{56}$Ni/$^{56}$Co in the
ejecta (see Fig.~\ref{fig:SNIa_spec}), probing the progenitors and
also testing the calibration of the cosmological distance
determination with SNe
Ia~\cite{1993ApJ...413L.105P}.
\g-rays from the $^{44}$Ti/$^{44}$Sc decay chain can be detected in
most of the young Galactic supernova remnants and in SN1987A,
providing information on the degree of asymmetry and clumpiness of the
ejecta of core-collapse
SNe~\cite{2018JHEAp..19....1D}.

\agam can detect line emission from kilonovae up to $12$ Mpc away and
measure the mass of nuclei synthesised by rapid neutron capture. In
the Milky Way, it can uncover the remnant of a kilonova up to
10--100~kyr old through long-lived radioisotopes such as $^{126}$Sn
and their fission
products~\cite{2019ApJ...880...23W,2020ApJ...889..168K,2020ApJ...903L...3W}.
\agam will map the positron annihilation line at 511 keV with
unprecedented quality and may even obtain the first
point-source detection.  It can detect the $2.2$-MeV neutron-capture
line from accreting neutron stars. It can also map the long-lived
$^{26}$Al and $^{60}$Fe radioisotopes, shedding new light on stellar
nucleosynthesis and on the subsequent mixing of the high-metallicity
ejecta into the
ISM~\cite{2021PASA...38...62D}. Finally,
it may provide the first unambiguous estimate of the $^{22}$Na and
$^7$Be
(see~ref.~\cite{2025A&A...698A.291I}
for INTEGRAL hints of the latter) yield in a nova, and thus gauge the
nova contribution to the Galactic enrichment in $^7$Li.

\vspace{0.4 cm}
\textbf{Cosmic Ray Sources and Feedback on Galaxy Evolution}
\vspace{0.1cm}

\agam data will be crucial to resolve important CR puzzles.
Observations of CR accelerators below 1 GeV will be decisive for
disentangling the emission of accelerated ions and electrons, and that
of freshly accelerated versus re-accelerated CRs, and will allow
studying CR feedback on the sources. \agam will provide the first
precise measurement of the ``Pion Bump'' below 500 MeV, a direct probe
of hadronic
emission~\cite{2024PhRvD.110f3026L}.
MeV-band spectra of CR sources will probe the initial acceleration of
CR electrons and hence the connection between CRs and quasi-thermal
plasma~\cite{2022RvMPP...6...29A},
while nuclear de-excitation lines will provide otherwise inaccessible
information on CR ions below $200$~MeV/nucleon.

The impact of CRs is a highly uncertain feedback factor in galaxy
evolution. The pressure gradient of GeV CRs can drive galactic
outflows, whereas low-energy ($<100$~MeV) CRs are the main source of
ionisation and heating in dark clouds, with direct impact on star
  formation. \agam observations will advance our poor understanding
of the propagation of these two important CR populations, probing
their spectra at different locations in the disk, near their sources,
in the ``quiet'' ISM and in the highly turbulent medium of starburst
regions, where re-acceleration can
occur~\cite{2011Sci...334.1103A,2020SSRv..216...42B}.
Mapping their large-scale distribution will measure beyond the
  Galactic disk the imprint of star-forming
regions~\cite{2015ARA&A..53..199G}.

In the Galactic center region, \agam will improve the spatial and
spectral characterisation of the puzzling \g-ray excess, to elucidate
its debated origin between DM annihilation, an excess of CRs, or a
population of unresolved millisecond
pulsars~\cite{2021PhRvD.103f3029D}.
On larger scales, the angular resolution and energy band of the
instrument are ideal to study the morphology and ion/lepton
composition of the Fermi bubbles, and help assess their origin as a
CR-driven wind or as jets powered by the central BH,
Sgr~A*~\cite{2014ApJ...793...64A,2022ApJ...927..225N}.

\vspace{0.4 cm}
\textbf{Deciphering the Energetic Transient Sky}
\vspace{0.1cm}

\begin{figure}[htbp]
\centering
\includegraphics[width=0.8\linewidth]{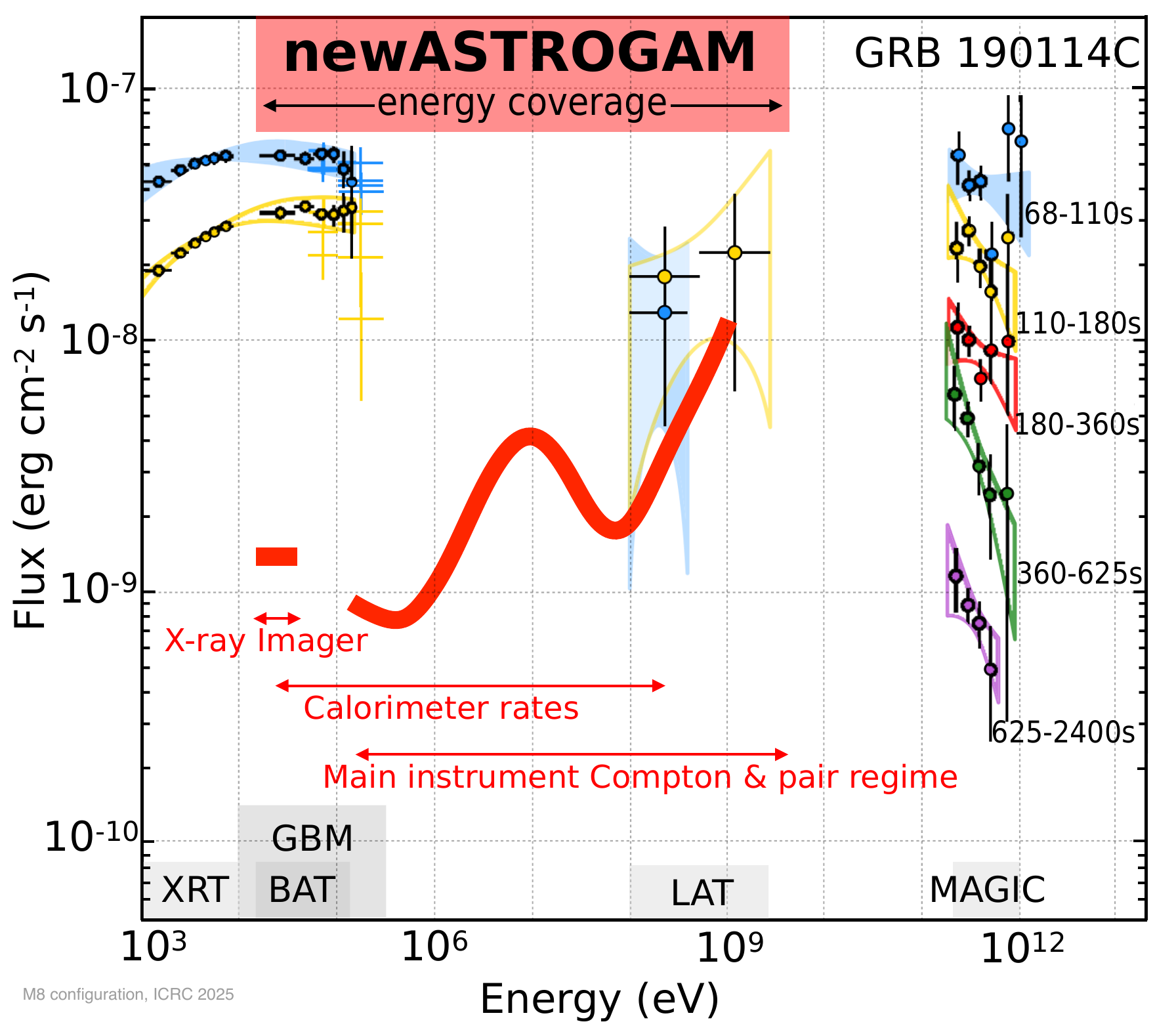}
\caption{X- and \g-ray afterglow spectrum of a long
  GRB~\cite{2019Natur.575..459M}. \agam
  probes the poorly covered transition region, its sensitivity is
  shown in red here for $600\,$s exposure. 
\label{fig:GRB}}
\end{figure}

The sensitivity, FoV, and spectral range of \agam will allow the
MeV/GeV-band detection of hundreds of long (see Fig.~\ref{fig:GRB})
and short GRBs. The latter, counterparts to binary NS mergers, will be
detected at a rate of several tens per year, ensuring synergy with GW
detectors. For the former, the mission will not only see the bright
prompt emission, but can uniquely follow the afterglow and determine
the radiation process and the spectrum of the radiating particles.

The X-ray Coded Mask will yield arcmin localisation, guiding
multi-wavelength follow-up observations of the source environment and
of the development of the kilonova as softer light emerges hours and
days later.  Moreover, \agam will measure intermediate X- to \g-rays
in a special burst event mode (\textit{calorimeter count rates}) and
provide polarimetry for the brightest events, key to identifying the
radiation process and constrain the magnetic and kinetic profile of
the
jet~\cite{2020LRR....23....4B}. Time-domain
surveys conducted by SKA in the radio, Roman in the infrared, and
Rubin in the optical, will revolutionise our knowledge of the
transient sky: \agam will add the fundamental MeV-GeV band to this
revolution, covering many source classes beyond the well-studied
micro-quasars, AGN and GRB transients, including
magnetars~\cite{2020Natur.587...59B}
and Fast Blue Optical
Transients~\cite{2019ApJ...872...18M}.
All detected transient signals will be rapidly disseminated with
arcminute localization in hard X-rays.

\vspace{0.4 cm}
\textbf{Fundamental and New Physics}
\vspace{0.1cm}

\agam will explore fundamental physics.  It will access the MeV to GeV
mass scale of DM, where direct searches are insensitive and collider
searches are highly model dependent, and offer multiple angles for
detections~\cite{2025PhRvD.111h3037O,2025arXiv250304907C}:
its angular resolution allows DM signals from subhalos and the inner
Galaxy to be better identified than with \textit{Fermi}, it offers
unprecedented details of DM models explaining the 511~keV
line~\cite{2024ApJ...973L...6D,2024arXiv241016379D},
and can differentiate pulsar and DM interpretations of the
Galactic-center
excess~\cite{2025PhRvD.111h3037O}. Weakly
Interacting Sub-eV Particles, e.g. axions and axion-like particles
(ALP)~\cite{1977PhRvL..38.1440P},
would produce spectral signatures in the MeV
band~\cite{2025arXiv250503038H}
or could be probed by ALP-induced GRBs from extragalactic
SNe~\cite{2010ARNPS..60..405J,2024PhRvD.110d3019L}.
Polarimetry of bright GRBs will address fundamental questions related
to vacuum birefringence and Lorentz-invariance
violation~\cite{2022ApJ...924L..29X}.
Finally, \agam will be highly sensitive to photon emission from
primordial BHs in the mass range $10^{15-17}$~g, which peaks at 1-30
MeV~\cite{2022PhRvD.106d3003K,2024PhRvD.109d3020X}.

\section{\agam Mission Configuration}
To achieve a breakthrough sensitivity, \agam will be launched in an equatorial low-Earth orbit of inclination $< 2.5^\circ$, eccentricity $< 0.01$, and altitude 550~--~600 km. Such an orbit is only marginally affected by the South Atlantic Anomaly therefore minimising instrumental particle backgrounds, ideal for high-energy observations~\cite{2022hxga.book...53T}. Alerts for new transient sources detected onboard with arcminute accuracy will be instantly transmitted to the ground. The mission will mostly do survey observations covering with its large FoV the entire sky in one day. In addition, occasional target-of-opportunity observations can be performed by a nearly inertial pointing to focus on a particular region. The nominal mission duration is planned to be three years.

The payload consists of a single instrument operating over five orders of magnitude in photon energy (15~keV--3~GeV), detecting photons via the photoelectric effect, Compton scattering, and e$^\pm$ pair production. To veto cosmic rays, the outermost Anti-Coincidence (AC) system is composed of segmented panels of plastic scintillators covering the top and the lateral sides of the roughly quadratic payload ($\approx$120~cm across). The AC vetoes more than 99.99\% of the penetrating charged particles and is based on heritage of the \textit{AGILE} and \textit{Fermi} missions. Within the AC and 14~cm before the first Silicon Tracker layer is a thin Coded Mask to image X-rays in the $\sim$15--40~keV energy range with an angular resolution of $\sim$15', providing a localisation capability of a few arcminutes for bright sources. The mask consists of a tungsten sheet with a pseudo-random pattern (open fraction 50\%). The mask mechanical support is also provided by an X-ray collimator of carbon fibre walls. The FoV of the X-ray coded mask monitor is 4.06~sr (half coded).

The instrument's heart is a Silicon microstrip Tracker. Such strips are already employed for \g-ray detection in space (e.g. for \textit{DAMPE}). The Tracker measures hard X-rays that have passed the Coded Mask and \g-rays that undergo Compton scattering or pair conversion. The baseline design has double-sided detectors (DSSDs) read out by low-noise and low-power electronics with self-triggering capability. The Tracker is supported by a very light carbon fibre mechanical structure. 9 DSSDs are wire bonded strip to strip to form a 2-D ladder. Alternatively, CMOS pixels~\cite{2024JInst..19C4010S,2022JATIS...8d4003C}
will be assessed and compared to the baseline DSSDs during the study phase.

The Calorimeter is a pixelated detector made of a high-$Z$ scintillation material (such as CsI(Tl)) for efficient absorption of \g-rays and e$^\pm$ pairs. It consists of an array of parallelepiped crystal bars read out by silicon drift detectors or silicon photomultipliers. The depth of interaction in each crystal is measured from the difference of recorded scintillation signals at both ends. The Calorimeter architecture is based on the heritage of instruments like for example \textit{INTEGRAL}/PICsIT or \textit{POLAR-2}.

\bibliographystyle{JHEP}
\bibliography{main_arXiv}

\newpage

\section*{\agam collaboration author list as submitted to ESA's M8 call (national contact points in bold)}
\thispagestyle{empty}

\noindent \textbf{Lead proposer ESA M8 step 1 proposal: David Berge}\\
\noindent Proposal co-PIs:
\noindent David Berge, M. Nicola Mazziotta, Uwe Oberlack, Vincent Tatischeff, Marco Tavani\\

\vskip 2mm
\noindent 
M.~Ackermann\textsuperscript{1},
M.~Ajello\textsuperscript{2},
R.~Aloisio\textsuperscript{3},
E.~Amato\textsuperscript{4},
G.~Ambrosi\textsuperscript{5},
A.~Argan\textsuperscript{6},
P.~von Ballmoos\textsuperscript{7},
F.~C.~T.~Barbato\textsuperscript{3},
M.~Barschke\textsuperscript{1},
R.~Battiston\textsuperscript{8}
% ...

\vskip 2mm
\noindent 
\textsuperscript{1}{\footnotesize Deutsches Elektronen-Synchrotron DESY Zeuthen, Germany};
\textsuperscript{2}{\footnotesize Department of Physics and Astronomy, Clemson University, Clemson, USA};
\textsuperscript{3}{\footnotesize Gran Sasso Science Institute and INFN LNGS, Italy};
\textsuperscript{4}{\footnotesize Istituto Nazionale di Astrofisica (INAF) OAA Arcetri, Italy};
\textsuperscript{5}{\footnotesize Istituto Nazionale di Fisica Nucleare (INFN) Perugia, Italy}; 
\textsuperscript{6}{\footnotesize INAF Headquarters Roma, Italy};
\textsuperscript{7}{\footnotesize IRAP Toulouse, France};
\textsuperscript{8}{\footnotesize University and INFN Trento, Italy}; 
\textsuperscript{9}{\footnotesize Humboldt University Berlin, Germany};
\textsuperscript{10}{\footnotesize University Mainz, Germany};
\textsuperscript{11}{\footnotesize Politecnico and INFN Bari, Italy};
\textsuperscript{12}{\footnotesize DTU Space, National Space Institute, Technical University of Denmark};
\textsuperscript{13}{\footnotesize SRON Netherlands Institute for Space Research, Netherlands}
\textsuperscript{14}{\footnotesize INAF-OAS Bologna, Italy};
\textsuperscript{15}{\footnotesize University W\"urzburg, Germany};
\textsuperscript{16}{\footnotesize INAF-IAPS Roma, Italy};
\textsuperscript{17}{\footnotesize NASA Goddard Space Flight Center, MD, USA};
\textsuperscript{18}{\footnotesize Dublin City University, Ireland};
\textsuperscript{19}{\footnotesize INFN Roma Tor Vergata, Italy};
\textsuperscript{20}{\footnotesize APC, Universit\'e de Paris, CNRS, France};
\textsuperscript{21}{\footnotesize LIP, Departamento de Fisica Universidade de Coimbra, Portugal};
\textsuperscript{22}{\footnotesize University and INFN Padova, Italy};
\textsuperscript{23}{\footnotesize INAF-IASF Palermo, Italy};
\textsuperscript{24}{\footnotesize INAF-OA Capodimonte, Italy};
\textsuperscript{25}{\footnotesize University of Paris-Saclay, CNRS, IJCLab, France};
\textsuperscript{26}{\footnotesize INFN Bari, Italy};
\textsuperscript{27}{\footnotesize LLR, \'Ecole Polytechnique, CNRS, Institut Polytechnique de Paris, France};
\textsuperscript{28}{\footnotesize University Bochum, Germany};
\textsuperscript{29}{\footnotesize University Erlangen, Germany};
\textsuperscript{30}{\footnotesize Institute of Cosmos Sciences of the University of Barcelona (ICCUB), Spain};
\textsuperscript{31}{\footnotesize University and INFN Perugia, Italy};
\textsuperscript{32}{\footnotesize INAF-OAB Merate, Italy};
\textsuperscript{33}{\footnotesize CEA/IRFU, Universit\'e Paris-Saclay, France};
\textsuperscript{34}{\footnotesize University College Dublin, Ireland};
\textsuperscript{35}{\footnotesize  ICE-CSIC and IEEC, Barcelona, Spain};
\textsuperscript{36}{\footnotesize Czech Technical University Prague, Czech Republic};
\textsuperscript{37}{\footnotesize Newcastle University, Great Britain};
\textsuperscript{38}{\footnotesize NTNU, Trondheim, Norway};
\textsuperscript{39}{\footnotesize Univ. Bordeaux, CNRS, LP2i Bordeaux, France};
\textsuperscript{40}{\footnotesize Physics Department, Stockholm University, Sweden};
\textsuperscript{41}{\footnotesize Department of Physics and Astronomy, University of Turku, Finland}; 
\textsuperscript{42}{\footnotesize University and INFN Trieste, Italy};
\textsuperscript{43}{\footnotesize University and INFN Bari, Italy};
\textsuperscript{44}{\footnotesize LUPM, Universit\'e Montpellier, CNRS, France};
\textsuperscript{45}{\footnotesize Space Research Center, Polish Academy of Sciences, Warszawa, Poland};
\textsuperscript{46}{\footnotesize Space Astronomy Lab, Nagoya University, Japan};
\textsuperscript{47}{\footnotesize University Darmstadt, Germany};
\textsuperscript{48}{\footnotesize Department of Astronomy, University of Geneva, Switzerland};
\textsuperscript{49}{\footnotesize INAF-OAR Roma, Italy};
\textsuperscript{50}{\footnotesize Royal Institute of Technology (KTH), Stockholm, Sweden};
\textsuperscript{51}{\footnotesize Karlsruhe Institute for Technology, Germany};
\textsuperscript{52}{\footnotesize University Potsdam, Germany};
\textsuperscript{53}{\footnotesize Institut de Física d'Altes Energies (IFAE), Barcelona, Spain};
\textsuperscript{54}{\footnotesize IAAT, University T\"ubingen, Germany};
\textsuperscript{55}{\footnotesize University of Amsterdam, Netherlands};
\textsuperscript{56}{\footnotesize Naval Research Laboratory Washington DC, USA};
\textsuperscript{57}{\footnotesize Department of Nuclear and Particle Physics, University of Geneva, Switzerland};
\textsuperscript{58}{\footnotesize Mullard Space Science Laboratory, University College, London, UK};
\textsuperscript{59}{\footnotesize Nicolaus Copernicus Astronomical Center, Polish Academy of Sciences, Warszawa, Poland};
\textsuperscript{60}{\footnotesize University of California at Berkeley, Space Sciences Laboratory, USA}.

\end{document}